\documentstyle[12pt,psfig]{article}

\newcommand{\rf}[1]{(\ref{#1})}
\newcommand{\bea}{\begin{eqnarray}}
\newcommand{\eea}{\end{eqnarray}}

\newcommand{\e}{{\rm e}}
\renewcommand{\d}{{\rm d}}

\renewcommand{\l}{\lambda}
\renewcommand{\L}{\Lambda}

\renewcommand{\a}{\alpha}

\newcommand{\m}{\mu}

\newcommand{\ep}{\varepsilon}

\newcommand{\del}{\delta}
\newcommand{\Del}{\Delta}

\newcommand{\oh}{\frac{1}{2}}
\newcommand{\oq}{\frac{1}{4}}

\newcommand{\tr}{{\rm Tr}}
\newcommand{\ra}{\right\rangle}
\newcommand{\la}{\left\langle}
\newcommand{\prt}{\partial}
\newcommand{\mi}{\!-\!}
\newcommand{\equ}{\!=\!}
\newcommand{\pl}{\!+\!}

\newcommand{\cD}{{\cal D}}

\newcommand{\cM}{{\cal M}}

\newcommand{\no}{\nonumber}
\newcommand{\nn}{\no\\}

\newcommand{\tdqg}{two-dimensional quantum gravity}

\newcommand{\ointz}{\oint \frac{dz}{2\pi i \, z}\;}

\newcommand{\SL}{\sqrt{\L}}
\newcommand{\SLT}{\sqrt{\L}T}

\def\void{}
\def\labelmark{}

\newenvironment{formula}[1]{\def\labelname{#1}
\ifx\void\labelname\def\junk{\begin{displaymath}}
\else\def\junk{\begin{equation}\label{\labelname}}\fi\junk}%
{\ifx\void\labelname\def\junk{\end{displaymath}}
\else\def\junk{\end{equation}}\fi\junk\labelmark\def\labelname{}}

{\ifx\void\labelname\def\junk{\end{array}\end{displaymath}}
\else\def\junk{\end{array}\right.\end{equation}}
\fi\junk\labelmark\def\labelname{}\def\junk{}
}

\newcommand{\beq}{\begin{formula}}
\newcommand{\eeq}{\end{formula}}
\newcommand{\beqv}{\begin{formula}{}}

\begin {document}
\hfill NBI-HE-98-31

\hfill AEI-080


\hfill  MPS-RR-1998-12

\begin{center}
\vspace{24pt}
{ \large \bf Euclidean and Lorentzian Quantum Gravity --\\
Lessons from Two Dimensions }

\vspace{24pt}

{\sl J.\ Ambj\o rn}$\,^{a,}$\footnote{email ambjorn@nbi.dk},
{\sl R.\ Loll}$\,^{b,}$\footnote{email loll@aei-potsdam.mpg.de},
{\sl J.L.\ Nielsen}$\,^{a,}$\footnote{email jlnielsen@nbi.dk} and
{\sl J.\ Rolf}$\,^{a,}$\footnote{email rolf@nbi.dk. Address after 
October 1st: Humboldt-Universit\"at, Berlin, Germany.}

\vspace{24pt}

$^a$~The Niels Bohr Institute, \\
Blegdamsvej 17, DK-2100 Copenhagen \O , Denmark 

\vspace{24pt}
$^b$~Max-Planck-Institut f\"{u}r Gravitationsphysik,\\
Albert-Einstein-Institut,\\
Schlaatzweg 1, D-14473 Potsdam, Germany

\vspace{36pt}

\end{center}

\vfill


\begin{center}
{\bf Abstract}
\end{center}

\vspace{12pt}
\noindent
No theory of four-dimensional quantum gravity exists as yet.
In this situation the two-dimensional theory, which can be analyzed 
by conventional field-theoretical methods, 
can serve as a toy model
for studying some aspects of quantum gra\-vi\-ty. It represents
one of the rare settings in a quantum-gravitational context where one can 
calculate quantities 
truly independent of any background geometry. 

We review recent 
progress in our understanding of 2d quantum gravity, and in
particular the
relation between the Euclidean and Lorentzian sectors of the
quantum theory. We show that conventional 2d Euclidean quantum 
gravity can be obtained from Lorentzian quantum gravity by
an analytic continuation only 
if we allow for spatial topology changes in the latter. 
Once this is done, 
one obtains a theory of quantum gravity where space-time is fractal:
the intrinsic Hausdorff dimension 
of usual 2d Euclidean quantum gravity is four, and not two. However,
certain aspects of quantum space-time remain two-dimensional,
exemplified by the fact that its so-called spectral 
dimension is equal to two.

\vfill

\newpage
\section{Introduction}

Quantum gravity plays havoc with our classical notions of probing the 
structure of space-time. In classical gravity we 
determine the metric structure of space-time either by using rods and
clocks or by studying 
the propagation of test particles, from which one can 
deduce a number of properties of the underlying geometry. 
When turning to {\it quantum} gravity, both methods encounter 
conceptual and technical difficulties since the object of interest
is no longer a single classical space-time, but a quantum average 
over all possible geometries.

At present there is no 
quantum gravity theory of four-dimensional Lorentzian 
metrics which could tell us how to probe the genuine quantum
structure of space-time.
String theory or recent suggestions of an even more 
general ``M-theory'' may eventually tell us how to determine
this quantum structure from first principles.
However, despite many grand claims they have until now 
taught us little beyond the standard picture of 
quantum fluctuations on classical background geometries. 
Attempts to go further and define genuinely new non-perturbative 
concepts replacing the classical notions of geometry are at best tentative.

It might thus be useful to try to address questions
about the quantum nature of space-time in two dimensions 
where a conventional quantum field theory of 
gravity {\it does} exist.
Such a theory contains no field-theoretical degrees of freedom
(it has no gravitons), but only a finite number of quantum-mechanical degrees 
of freedom. Nevertheless many 
of the conceptual problems occurring in four dimensions
are already present in the two-dimensional theory: 
how do we define observables 
invariant under reparametrization? How do we define the propagation
of particles? How do we determine simple properties of 
space-time, in absence of a fixed reference geometry? 
In a way the quantum nature of these questions is at least as 
pronounced in 2d as it is in 4d,
despite the fact that the action is trivial and non-dynamical. 
Namely, the triviality of the 
action implies that all 2d geometries have to be included with 
the same weight in the quantum average. The concept of a 
fixed background geometry around which one can study small perturbations
therefore simply does not exist.

Another moot point in studies of quantum gravity is the question of 
the signature of the space-time metric.
One might hope for the existence of a theory in which 
the signature was determined dynamically.
Even if we are less ambitious and consider the signature as
given {\it a priori}, nothing much is known 
about the relation between the 
quantum theories with Lorentzian and Euclidean signatures.
Already classically the simple expedient of
applying a Wick rotation $t\rightarrow \tau=it$ fails in all
but a few special cases, for instance, when the space-time is
static and thus admits a {\it global} choice of time $x_0$ such that 
the cross terms $g_{0i}$, $i\geq 1$ of the metric tensor vanish and
its spatial components $g_{ij}$ are time-independent. 
We do not know of a way to set up a 1-to-1 correspondence between
generic solutions of the continuum Einstein equations with different
signatures. If it exists, it is likely to be technically
involved. From this point of view it is unclear if one 
can expect to obtain the correct Lorentzian theory 
by first performing a path integral over general Euclidean metric 
configurations and then analytically continuing in some 
way, as is the case in conventional field theory in 
flat space-time. Again, two-dimensional quantum gravity 
is a good laboratory for the study of such questions,
since the theory can be treated by conventional field-theoretical tools,
and turns out to be non-perturbatively renormalizable.

In the next section we briefly describe a quantum gravity 
analogue of the classical rods-and-clocks measurement, that is, how
to extract diffeomorphism-invariant information about the 
geometry of quantum space-time.
Section 3 contains some details of
a discussion of test particle propagation in quantum space-time.
We then define a path-integral quantization of metric space-times
with Lorentzian signature and show that the fractal structure 
of space-time only agrees with the one obtained in Euclidean 
gravity if topology changes of {\it space} are included.  
The final section contains a discussion of the results and  
some remarks on possible generalizations to higher dimensions.

\section{The functional integral over 2d geometries}

The partition function for two-dimensional Euclidean 
gravity is given by
\beq{8h}
Z(\L)= \int \cD [g] \; \e^{-\L V_g},~~~~~~~V_g \equiv \int_{\cM} \d^2\xi 
\sqrt{\det g(\xi)},
\eeq
where $\cD [g]$ denotes a suitable measure on the space of 
equivalence classes $[g]$ of Riemannian two-metrics, and $\L$ is the
bare cosmological constant. The weight factor 
in the exponential is proportional to the Einstein action, which in
the present case consists only of the cosmological term. (The scalar 
curvature term has been omitted since it gives only a constant contribution 
for fixed space-time topology.)
It is sometimes convenient to consider the partition function 
where the volume $V$ of space-time is kept fixed. We define it by
\beq{9h}
Z(V) = \int \cD [g] \; \del (V-V_g),
\eeq
such that 
\beq{10h}
Z(\L) = \int_0^\infty \d V \; \e^{-V\L} Z(V).
\eeq

In the following we will discuss a set of reparametrization-invariant 
``obser\-vables", namely, the Hartle-Hawking
wave functionals and the two-point functions of the theory. 
The Hartle-Hawking wave functional is defined by
\beq{50}
W_\L(L) = \int \cD [g] \; \e^{-S(g;\L)},
\eeq
where $L$ stands for the metric degrees of freedom 
on the {\em boundary} $\partial\cM$ of the ma\-ni\-fold $\cM$. 
In 2d the equivalence class of the 
boundary metric (in the case of one boundary component) 
is uniquely fixed by its length, which we will again denote by $L$. 
In dimensions higher than two, instead of the
single parameter $L$ one would need to specify  
a metric equivalence class $[g|_{\partial\cM}]$ on $\partial\cM$. 
The functional 
integration would then have to be performed over all metric
configurations in the
interior which induce $[g|_{\partial\cM}]$ on the boundary.

It is often convenient to consider 
boundaries with variable length $L$ by introdu\-cing a {\em boundary
cosmological term} in the action, resulting in
\beq{51}
S(g;\L,X) = \L \int_{\cM} \d^2\xi \sqrt{\det g(\xi)} + X \int_{\prt \cM} \d s,
\eeq  
where $\d s$ is the invariant line element induced by $g$ 
on the boundary and $X$ is called the boundary cosmological constant.
We may then define
\beq{52}
W_\L(X) = \int \cD [g]\; \e^{-S(g;\L,X)},
\eeq
so that the wave-functions $W_\L(L)$ and $W_\L(X)$ are related by a Laplace
transformation in the boundary length,
\beq{53}
W_\L(X) = \int_0^\infty \d L \; \e^{-X L} \, W_\L(L).
\eeq

The two-point function $G_\L(R)$ is defined by 
\beq{53h} 
G_\L(R) = \int \cD [g] \; \e^{-S(g,\L)} 
\int\!\!\! \int \d^2\xi\sqrt{\det g(\xi)} \, \d^2 \eta \sqrt{\det g(\eta)} 
\; \del(D_g (\xi,\eta)-R),
\eeq
where $D_g(\xi,\eta)$ denotes the geodesic distance between 
the points $\xi$ and $\eta$
in the given metric $g_{ab}$. Again, it is sometimes convenient to consider 
a situation where the space-time volume $V$ is fixed. 
The corresponding function
$G_V(R)$ will be related to \rf{53h} by a Laplace transformation, analogous 
to the partition functions $Z$ above,
\beq{54}
G_\L(R) = \int_0^\infty \d V \; \e^{-V\L}\, G_V(R).
\eeq
Both $G_\L(R)$ and $G_V(R)$ have the interpretation of 
partition functions for universes with pairs of marked points separated by
a given geodesic distance $R$. If we denote the average 
volume of a spherical shell of geodesic radius $R$ in the class 
of metrics with space-time volume $V$ by  $S_V(R)$, we have by definition
\beq{55}
S_V(R) = \frac{G_V(R)}{V Z(V)}.
\eeq
{\it The function $S_V(R)$ therefore 
serves as the quantum analogue of the volume of a
classical spherical shell of radius $R$ in a fixed geometry.} In this 
sense it is the quantum version of a ``rods-and-clocks'' measurement 
of the geometry. In the same spirit one may define an intrinsic 
fractal dimension $d_H$ of the ensemble of
metrics by the scaling behaviour
\beq{55a}
\lim_{R \to 0} S_V(R) \sim R^{d_H-1}( 1 + O(R)). 
\eeq 
Alternatively, one could introduce another notion of
dimensionality $d_H$ via
\beq{55b}
\la V \ra_R \sim - \frac{\prt \, \log G_\L(R)}{\prt \L} \sim R^{d_H},
\eeq
for a suitable range of $R$ related to the value of $\L$. As will
become clear, 
these two definitions agree in the case of pure gravity.  Eq.\ \rf{55a}
can be viewed as a ``local'' definition of $d_H$, 
while eq.\ \rf{55b} characterizes a  ``global'' property of space-time.  
Since the two definitions result in the 
same value for $d_H$, two-dimensional gravity 
can be said to have a genuine fractal dimension over all scales.

In the regularization we will be using each discrete geometry,
representing an equivalence class of metrics $[g]$, will be included
with the same weight. The functional integral $\int \cD [g]$ is turned
into a discrete sum over inequivalent geometries, each with weight one.
Eq.\ \rf{9h} then entails that the calculation of $Z(V)$ basically 
amounts to a counting problem. 
The same is true for the other observables defined above. One way of
performing the summations is to introduce a suitable regularization of the 
set of geometries by means of a cut-off, perform the summation, 
and then to remove the cut-off.

\subsection{The regularization}

In the calculations referred to below we regularize the 
integral over geometries by using equilateral triangles as elementary
building blocks, which are glued 
together in all possible ways consistent with a given space-time topology  
{\cite{david1,adf,kazakov,kkm}}. The metric degrees of freedom
are encoded in the different ways of gluing the triangles. For example,
the deficit angle at a vertex $v$ of the triangulation (a measure
of local curvature) is determined by the number of triangles meeting at $v$.
From this point of view a summation over such triangulations
forms a grid in the class of Riemannian 
geometries associated with a given manifold $\cM$. The hope is 
that in the limit as the edge-length $a$ of the triangles is
taken to zero, 
this grid becomes sufficiently dense and uniform to approximate 
correctly the functional integral over all Riemannian geometries.
We will provide evidence that this is the case by explicit
calculations that are in agreement with the corresponding continuum 
expressions, where they can be compared.

The surprising situation encountered in two-dimensional
quantum gravity is the fact that the analytical power of the regularized 
theory seems to exceed that of the formal continuum manipulations 
performed within Liouville theory. This reverses the usual situation
where regularized theories are either used 
in a perturbative context to remove infinities order by order, or introduced  
in a non-perturbative setting to make numerical simulations possible.
By contrast, in the regularized framework of 2d quantum gravity at hand,
we will derive analytic (continuum) expressions with an ease which can 
presently not be matched by formal continuum manipulations.

\subsection{The two-loop and two-point functions}

We first define a generalization of the two-point
function: consider universes with the topology of a finite 
cylinder, where the exit loop (the outgoing $S^1$-boundary
component) is separated from the entrance loop (the incoming
$S^1$-boundary) by a geodesic distance $R$. 
By definition this means that each point on the exit loop 
has a geodesic distance $R$ to the entrance loop (but not 
necessarily the converse). The partition 
function for such universes will be denoted by
\beq{*l1}
G_\L(L_1,L_2;R),
\eeq
where $L_1$ and $L_2$ are the lengths of the 
entrance and exit loops. 
Introducing boun\-dary cosmological 
constants $X$ and $Y$ for the entrance and exit loops we obtain 
\beq{*l2}
G_\L(X,Y;R) = \int_0^\infty dL_1 \int_0^\infty dL_2\; 
\e^{-L_1X}\e^{-L_2Y}\;G_\L(L_1,L_2;R)
\eeq 
by a Laplace transformation.

In section \ref{causal} we will describe our regularized model in
some detail, but for the moment we are only interested in some
general aspects independent of those details.
In a regularized model we must introduce dimensionless
coupling constants $\l_{b_1}$, $\l_{b_2}$   
and $\l$ corresponding to the bare couplings $X$, $Y$ and $\L$. 
Since $X$ ($Y$) and $\L$ 
have dimensions of mass and (mass)$^2$, we expect additive
renormalizations of the form 
\beq{*l3}
\l = \l_c + a^2 \L,~~~\l_{b_1} = \l_{b}^c + a X,
~~~\l_{b_2} = \l_{b}^c +a Y.
\eeq
For convenience we introduce the notation 
\beq{*l3a}
g=\e^{\mi \l}=g_c(1\mi \oh a^2\L),~~~~x= \e^{\mi\l_{b_1}} = x_c(1\mi a X),~~~~
y=\e^{\mi\l_{b_2}} = x_c(1\mi a Y).
\eeq
Strictly speaking this involves an analytic coupling constant redefinition 
of $X,Y$, compared to \rf{*l3}.  
In this way we obtain regularized two-loop functions
$G_\l(x,y;r)$ and $G_\l(l_1,l_2;r)$, where the integer $r$ is the 
geodesic distance measured in lattice units, and $l_1,l_2$ are the lengths
of the entrance and exit loops, again measured in lattice units.
$G_\l(x,y;r)$ will be related to $G_\l(l_1,l_2;r)$ 
by the discretized analogue of \rf{*l2}, namely, 
\beq{9}
G_\l(x,y;r)\equiv \sum_{k,l} x^k\,y^l \;G_\l(k,l;r),
\eeq
illustrating that from a combinatorial point of view 
$G_\l(x,y;r)$ can be regarded as the generating functional 
for the numbers $G_\l(l_1,l_2;r)$.
 
The two-loop function satisfies the combinatorial identity 
\beq{7}
G_\l(l_1,l_2,r_1+r_2) = \sum_{l} G_\l(l_1,l;r_1)\; G_\l(l,l_2;r_2),
\eeq
and in particular
\beq{8}
G_\l(l_1,l_2;r+1) =\sum_{l} G_{\l}(l_1,l;1)\;G_\l(l,l_2;r).
\eeq
The function $G_\l(l_1,l_2;1)$ appears on the right-hand side
as a transfer matrix, and 
knowing $G_\l(l_1,l_2;1)$ allows us to find $G_\l(l_1,l_2;r)$
by iterating \rf{8} $r$ times. As usual in combinatorics, this program 
is conveniently carried out by means of the generating function 
$G_\l(x,y;r)$ for the numbers
$G_\l(l_1,l_2;r)$. We rewrite relation \rf{7} as
\beq{10}
G_\l(x,y;r_1+r_2) = \ointz G_\l(x,z^{-1};r_1) G_\l(z,y;r_2),
\eeq
where the contour should be chosen to include the singularities 
in the complex $z$--plane of $G_\l(x,z^{-1};r_1)$ but not those
of $G_\l(z,y;r_2)$. 

In the same way we can define the regularized Hartle-Hawking wave functionals
$w_\l(l)$ and $w_\l(x)$ corresponding to $W_\L(L)$ and $W_\L(X)$. 
The former, $w_\l(l)$,  
will be a sum over an appropriate class of triangulations
with the topology of a disc and a boundary of lattice 
length $l$, and the continuum relation \rf{53}
replaced by 
\beq{53a}
w_\l(x) = \sum_l x^l \, w_\l(l).
\eeq
While the scaling of $G_\l(x,y;r)$,
\beq{*l10}
G_\l(x,y;r) \to a^{-1} G_\L(X,Y;R)~~~\mbox{for}~~~a \to 0,
\eeq
follows directly from \rf{10} and \rf{*l3}, 
we will assume that $w_\l(x)$ has the general form 
\beq{an3}
w_\l(x)= w^{ns}_\l(x)+ a^{\eta}W_\L(X) + \mbox{less singular terms}.
\eeq

\begin{figure}
\centerline{\hbox{\psfig{figure=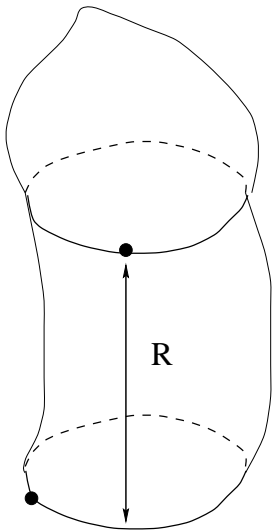,height=7cm,angle=0}}}
\caption[identity]{Marking a vertex in the bulk of $W_\L(X)$. The vertex
has a distance $R$ from the boundary loop, which has one marked vertex}
\label{identity}
\end{figure}

We now introduce an explicit mark in the bulk of $w_\l(x)$
by differentiating with respect to $\l$. This leads to the
combinatorial identity 
\beq{an1}
\frac{\prt w_\l(x)}{\prt \l} =
\sum_r \sum_l G_\l(x,l;g;t) \, l w_\l(l),
\eeq
which after a Laplace transformation becomes
\beq{an2}
\frac{\prt w_\l(x)}{\prt \l}=
\sum_t \ointz G_\l(x,z^{-1};r)\; \frac{\prt w_\l(z)}{\prt z}.
\eeq 
This is illustrated in fig.\ \ref{identity}. 
A given mark  has a 
distance $r$ ($R$ in the continuum) to the entrance loop. 
In the figure we have drawn all points with the same distance 
to the entrance loop and which form a connected loop containing the 
marked point. 
Let us now assume a general scaling of the form 
\beq{an3a}
R = a^{\ep} r,~~~~~\ep >0,
\eeq
for the geodesic distance $R$ in the continuum limit. 
Inserting eqs.\ \rf{an2} and \rf{an3a} into \rf{an1} we obtain
\beq{an4}
\frac{\prt w^{ns}_\l}{\prt \l}- a^{\eta-2} 
\frac{\prt W_\L(X)}{\prt \L} = \frac{1}{a^{\ep}} 
  \int d R \int dZ\;  G_\L(X,-Z;R) 
\bigg[ \frac{\prt w^{ns}_\l}{\prt z} -a^{\eta-1} \frac{\prt W_\L(Z)}{\prt Z}
\bigg],
\eeq
where $(x,\l)=(x_c,\l_c)$ in the non-singular part. 

It follows from eq.\ \rf{an4} that the only consistent choices 
for $\eta$ are
either $\eta < 0$ (in which case $\ep =1$) or $1< \eta < 2$. We will
presently consider the latter situation, and later in section \ref{causal}
obtain an interpretation for the case $\eta < 0$, $\ep =1$.   
For $1< \eta < 2$, eq.\ \rf{an4} splits into the two relations 
\beq{an5}
-a^{\eta-2} \frac{\prt W_\L(X)}{\prt \L} = \frac{1}{a^{\ep}}\, 
\frac{\prt w_{ns}}{\prt z}\bigg|_{z=x_c}\;
  \int d R \int dZ\;  G_\L(X,-Z;R) 
\eeq  
and 
\beq{an6}
\frac{\prt w_{ns}}{\prt g}\bigg|_{g=g_c} = -\frac{a^{\eta-1}}{a^{\ep}}  
  \int d R \int dZ\;  G_\L(X,-Z;R) 
\;\frac{\prt W_\L(Z)}{\prt Z}.
\eeq
{\it We are led to the conclusion that $\ep= 1/2$ and $\eta=3/2$.
This implies that the geodesic distance $R$ has 
an anomalous dimension.}
Note for future reference that eq.\ \rf{an5} is of the form
\beq{an5a}
-\frac{\prt W_\L(X)}{\prt \L} = \mbox{const.}\;G_\L(X,L_2=0),
\eeq
where $G_\L(L_1,L_2)$ is a (generalized) Hartle-Hawking 
wave functional with two boun\-daries, the cylinder amplitude
\beq{*l20}
G_\L(L_1,L_2) = \int_0^\infty dR\; G_\L(L_1,L_2;R).
\eeq

The disc amplitude $W_\L(X)$ can be calculated explicitly
using either combinatorial techniques 
(see \cite{book1} for a review) or matrix model methods (\cite{david,ajm}), 
and is found to have a 
dimension corresponding to $\eta = 3/2$.
Above we have reproduced this non-trivial result from 
simple combinatorial arguments. These can be taken further within
the dynamical triangulations approach to calculate explicitly 
the transfer matrix
$G_\l(x,y;1)$ (see \cite{japanese} or \cite{book1} for details), 
which appears in eq.\ \rf{10} in the special case
when $r_1 =1$,
\beq{10a}
G_\l(x,y;r+1) = \ointz G_\l(x,z^{-1};1) G_\l(z,y;r).
\eeq
An even simpler (and at the same time more general) derivation 
of the corresponding continuum equation 
will be given in section \ref{causal} below, in the context of
a slightly different model. Since two-dimensional 
Euclidean quantum gravity corresponds to $\eta=3/2$, $\ep=1/2$, 
we can substitute these values into 
eq.\ \rf{top13} below to derive the equation 
\beq{*l30}
\frac{\prt}{\prt R} G_\L(X,L_2;R) = -\frac{\prt}{\prt X}
\Big[ W_\L(X) \, G_\L(X,L_2;R) \Big]
\eeq
for $G_\L(X,L_2,R)$
(which also could have been obtained directly from \rf{10a}).
Eq.\ \rf{*l30} determines the structure of two-dimensional
Euclidean quantum gravity. First, it allows us to calculate
$W_\L(X)$, the Hartle-Hawking wave functional. Integrating 
\rf{*l30} with respect to $R$ and using relation \rf{an5a}
as well as the boundary condition $G_\L(X,L_2=0;R=0)=1$ 
(originating from $G_\L(L_1,L_2;R=0) = \delta(L_1-L_2)$),
we obtain
\beq{top18}
-1 = \frac{\prt}{\prt X}\bigg[ W_\L(X) \frac{\prt}{\prt\L} W_\L(X)
\bigg].
\eeq
Since $W_\L(X)$ has length dimension --3/2 ($\eta = 3/2$), i.e. 
$W_\L^2(X) = X^3 F(\SL/X)$, the general solution must be of the form 
\beq{top19}
W_\L(X) = \sqrt{-2\L X + b^2 X^3+ c^2 \L^{3/2}}.
\eeq
From the very origin of $W_\L(X)$ as the Laplace transform of the  
Hartle-Hawking wave functional $W_\L(L)$ which is bounded, 
it follows that $W_\L(X)$ has 
no singularities or cuts for $\mbox{Re}\, X >0$. This leads, 
after a suitable rescaling of the cosmological constant $X$,
to the expression
\beq{top20a}
W_\L(X) = \Big(X- \oh \SL\Big)
\sqrt{X+\SL},
\eeq
well-known from much more 
complicated calculations using matrix models and 
dynamical triangulations.
Once we know $W_\L(X)$, we can solve eq.\ \rf{*l30}
and obtain $G_\L(R)$ as the integral
\beq{*l31}
G_\L(R) = \int \d L_2 \;G(L_1 \equ 0,L_2;R) \, L_2 W_\L(L_2).
\eeq
This combinatorial identity is again illustrated in fig.\ 
\ref{identity}. Contracting $L_1$ to zero corresponds to 
introducing a marked point. Let the other marked point be 
at geodesic distance $R$. It will be located at a loop of some 
length $L_2$, having a distance $R$ from the first marked point. All such  
manifolds are obtained by closing the loop with length $L_2$ by 
a ``cap'' of length $L_2$ and integrating over all $L_2$. This 
is precisely the combinatorial content of expression 
\rf{*l31}. By a direct calculation one 
obtains
\beq{82}
G_\L(R) = \L^{3/4} 
\frac{\cosh \Bigl[R \sqrt[4]{\L}\Bigr]}{\sinh^3 \Bigl[ R
\sqrt[4]{\L}\Bigl]}.
\eeq
Using this expression it follows that  
\beq{newg1}
\la V \ra_R \sim  - \frac{\prt \, \log G_\L(R)}{\prt \L} \sim R^4
(1+ O(R^4\L)),
\eeq
implying that $d_H=4$ rather than 2, as one naively would have expected.
Observe also that $G_\L(R)$ falls off exponentially for $R \gg \sqrt[4]{\L}$.
It has been shown that this is a consequence of the same kind of 
sub-additivity arguments which lead to the exponential fall-off
of the massive particle propagator \cite{aw}. 

The quantity $d_H$ above is the ``globally defined'' 
Hausdorff dimension in the 
sense discussed below \rf{55b}, as is clear from eq.\ \rf{82}.
We can determine the ``local'' dimension $d_H$, defined by eq.\ \rf{55a},
by performing the inverse Laplace transformation
of $G_\L(R)$ to obtain $G_V(R)$. The average volume $S_V(R)$ of a 
spherical shell of geodesic radius $R$ in the ensemble of universes
with space-time volume $V$ can then be calculated from relation
\rf{55}, yielding
\beq{85}
S_V(R) = R^3 F(R/V^{\oq}),~~~~F(0) > 0,
\eeq   
where $F(x)$ can be expressed in terms of certain generalized hypergeometric
functions {\cite{ajw}}. Eq.\ \rf{85} shows that also the 
``local'' dimension $d_H$ is equal to four.

\subsection{Summary}

It has been shown how one can calculate the functional 
integral over two-dimensional geometries, by proceeding 
in close analogy with the 
functional integral over random paths. One of the fundamental  
results of the latter theory is that a 
generic random path between two points         
in $R^d$, separated a geodesic distance $R$, is {\em not} proportional to 
$R$ but to $R^2$.
This famous result has a direct translation to the theory of random 
two-dimensional geometries: the generic volume of a closed universe 
of radius $R$ is {\em not} proportional to $R^2$ but to $R^4$.

\section{The spectral dimension of space-time}

In the preceding section we have described a 
quantum version of the classical measurements of  
space-time geometry.
We now turn to the test particle aspect.
Consider the propagation of test particles in a fixed geometry.
By observation we can determine their propagator, and by an 
inverse Laplace transformation the associated 
heat kernel (see below). The heat kernel has an asymptotic 
expansion whose individual terms are functions of 
various contractions of the Riemann tensor  
and other geome\-tric quantities, and thus 
have a direct geometric interpretation.
One may wonder how much of 
this information survives in a genuine theory of quantum gravity 
where we integrate over {\it all} geometries. 
Such a study leads naturally to the 
notion of {\it spectral dimension} as a measure of 
the structure of quantum space-time. 

The most intuitive definition of the spectral dimension 
is based on the diffusion equation on a (compact) manifold with 
metric $g_{ab}$. Let $\Del_g$ denote the Laplace-Beltrami 
operator corresponding to $g_{ab}$. The probability distribution
$K(\xi,\xi';\tau)$ of diffusion in a fictitious ``time'' $\tau$ 
is related to the massless scalar propagator 
$(-\Del_g)^{-1}$ by \footnote{Since we consider compact manifolds, 
the Laplace-Beltrami operator $\Del_g$ has zero modes. Eq.\ \rf{1.1}
should be understood with these zero modes projected out. This is indicated 
by the prime.}
\beq{1.1}
\la \xi'\Big| (-\Del_g)^{-1} \Big| \xi\ra' 
= \int_0^\infty\! \d \tau \; K'(\xi,\xi';\tau) .
\eeq
In particular, the average return probability distribution at ``time'' $\tau$ 
has the small-$\tau$ behaviour
\beq{1.2}
RP'_g(\tau) \equiv \frac{1}{V_g}\int\! \d^d \xi \sqrt{\det g} \; 
K'(\xi,\xi;\tau) 
\sim \frac{1}{\tau^{d/2}}(1+O(\tau)),
\eeq
where $V_g$ denotes the volume of the compact manifold with metric $g$.
The important point in the context of quantum gravity is that $RP'_g(\tau)$ 
is invariant under reparametrization. We may therefore define
the quantum average over geometries,
\beq{1.2a}
RP'_V(\tau) \equiv \frac{1}{Z_V}  \int\! \cD [g]_V\;\e^{-S_{\rm eff}([g])} 
RP'_g(\tau),
\eeq
where  $S_{\rm eff}([g])$ denotes the effective
action of quantum gravity after the integration of possible matter
fields, and $Z_V$ denotes the partition function of quantum gravity plus 
matter for a fixed space-time volume $V$. 
The {\it spectral} dimension $d_s$ in quantum gravity is 
now defined by the small-$\tau$ behaviour of the functional average 
$RP'_V(\tau)$, 
\beq{1.2b}
RP'_V(\tau) \sim \frac{1}{\tau^{d_s/2}}(1+ O(\tau)).
\eeq
The $O(\tau)$ term in \rf{1.2} has a well-known asymptotic expansion 
in powers of $\tau$, where the coefficient of $\tau^r$ is an integral 
over certain powers and contractions of the curvature tensor.
This asymptotic expansion breaks down when $\tau \sim V^{2/d}$ at which 
point the exponential decay in $\tau$ of the heat kernel $K$ takes over.
If we average over all geometries as in \rf{1.2a}, it is natural to 
expect that the only invariant left will be the volume $V$ which is kept
fixed. Thus we expect a relation of the form
\beq{1.3}
RP'_V(\tau) = \frac{1}{\tau^{d_s/2}} F\Big(\frac{\tau}{V^{2/d_s}}\Big),
\eeq  
where $F(0) > 0$ and $F(x)$ falls off exponentially fast for
$x \to \infty$. 

For a fixed manifold of dimension $d$ and a given smooth geometry $[g]$ 
we have $d=d_s$ by definition. The functional average can in principle 
change this value, i.e.\ the dimension of $\tau$ may become anomalous. 
A well-known example of a similar nature can be found 
for the ordinary free particle. In the path-integral representation
of the free particle, any smooth path has of course fractal dimension
one. Nevertheless the short-distance properties of the free 
particle reflect the fact that the generic path contributing to the 
path integral has fractal dimension (the {\it extrinsic} 
Hausdorff dimension in the target space $R^D$) 
$D_H=2$ with probability one. In the same way the functional integral over 
geometries might change $d_s$ from the ``naive'' value $d$.
In \tdqg\ it is known, as mentioned above, 
that the intrinsic Hausdorff dimension {\it is} different from $d=2$
and the generic geometry is in this sense fractal, with probability one.
When one considers diffusion on fixed fractal structures 
(often embedded in $R^D$), it is well-known 
that $d_s$ can be different from both $D$ and the fractal dimension $d_h$
of the structure. If $\del$ denotes the so-called anomalous gap exponent,
defined by the relation between the diffusion time $\tau$ and 
the average spread of diffusion on the fractal structure, but measured in 
$R^D$,
\beq{1.2c}
\la r^2(\tau)\ra \sim \tau^{2/\del},
\eeq
the relation between the fractal dimension ({\em intrinsic}
Hausdorff dimension) $d_h$ of the structure,
the spectral dimension of the diffusion and the gap exponent is given by
\beq{1.2d}
d_s = \frac{2d_h}{\del}.
\eeq
If $\del$ is not anomalous, i.e.\ $\del =2$ as for diffusion on a smooth 
geometry, we have $d_s = d_h$, which is the analogue of $d_s =d$ for 
fixed smooth geometries. However, in general $\del \neq 2$ (for a review
of diffusion on fractal structure, see, for example, \cite{review}).

\subsection{The spectral dimension of 2d quantum gravity}

Let us now present a proof that $d_s=2$ in two-dimensional 
Euclidean gravity coupled to conformal matter with a 
central change $c \leq 1$ (i.e.\ for the range of conformal 
theories where Liouville field theory is applicable to 
2d quantum gravity coupled to matter).

Since $\Del_{\l g}= \l^{-1} \Del_g$, we have from definition
\rf{1.1} that 
\beq{jac1}
\tr' \Big(-\Del_{\l g}\Big)^{-1} = \l \tr' \Big(-\Del_{g}\Big)^{-1}.
\eeq
According to Liouville theory applied to two-dimensional 
quantum gravity coupled to a conformal field theory with 
central charge $c$, a general spin-less operator, $\Phi_n[g]$, 
which depends only on
the metric $g_{ab}$, which is diffeomorphism-invariant, and 
which satisfies
$\Phi[\l g] = \l^{-n}\Phi_n[g]$ at the classical level,
has the following scaling of its quantum expectation value\footnote{
The expectation value $\la \Phi_n [g]\ra_V$ 
of this operator in the context of \tdqg\
coupled to a conformal field theory with central charge $c$
is defined by
$$
\la \Phi_n [g]\ra_V = \frac{1}{Z_V} \int\cD [g]_V \e^{-S_{eff}[g]} \Phi_n[g],
$$
where $e^{-S_{eff}[g]}$ again denotes the effective action 
after integration over the matter fields, and  $Z_V[g]$ denotes the 
corresponding partition function for fixed space-time volume $V$.}
(see \cite{watabiki} for details):
\beq{2.2a}
\la \Phi_n[g]\ra_{\l V} = \l^{\a_{-n}/\a_1} \la \Phi_n[g]\ra_V,~~~~~~
\a_n= \frac{2n\sqrt{25-c}}{\sqrt{25-c}+\sqrt{25-c -24n}}.
\eeq
It follows that the quantity
\beqv
\frac{1}{V}\la\tr'\Big(-\Del_g\Big)^{-1}\ra_V
\eeq
is invariant under the transformation $V \to \l V$.
On the other hand, the assumed scaling \rf{1.3} implies that
\bea
\frac{1}{V}\la\tr'\Big(-\Del_g\Big)^{-1}\ra_V &=& 
\int_0^\infty \d \tau \; RP'(\tau) 
= \int_0^\infty  \d \tau
\frac{1}{\tau^{d_s/2}} F\Big(\frac{\tau}{V^{2/d_s}}\Big)\nonumber\\
&=&
\mbox{const.}\times V^{2/d_s-1}. \label{2.2b}
\eea
Thus it follows that $d_s=2$.

To corroborate this result, let us use the same kind of argument 
to derive a simple relation between the spectral dimension 
and the extrinsic Hausdorff dimension for dynamical self-similar 
systems like random surface models,  
branched polymers etc. (see \cite{abnrw} for details, for a 
definition of branched polymer models see \cite{adfo,adj}). 
Since the extrinsic 
Hausdorff dimension is known for these systems it will allow a
determination of the spectral dimension. As a typical example 
of such models (and maybe the most interesting), we consider random 
surfaces, here described as two-dimensional quantum gravity 
coupled to $D$ Gaussian fields $X_\m$. These fields can be 
viewed as embedding coordinates for a surface in $R^D$.
The partition function for such a system is given by 
\beq{2.1}
Z_V = \int\! \cD [g]_V \cD [X_\m]_{\rm cm}\, 
\e^{-\!\int\! \d^2 \xi \sqrt{\det g} \, g^{ab} \prt_a X_\m \prt_b X_\m}.
\eeq
Here $\cD [X_\m]_{\rm cm}$ denotes the functional 
integration over the $D$ Gaussian fields $X_\m$, but with the center of 
mass fixed (to zero). The extrinsic Hausdorff 
dimension $D_H$ is usually defined as 
\beq{2.2}
\la X^2 \ra_V \sim V^{2/D_H}~~~~{\rm for}~~~V \to \infty,
\eeq
where
\beq{2.3}
\la X^2 \ra_V \equiv \frac{1}{Z_V} \int\! \cD [g]_V \cD [X_\m]_{\rm cm}\,
\e^{-\!\int\! \d^2 \xi \sqrt{\det g} \, g^{ab} \prt_a X_\m \prt_b X_\m}\;
\left(\frac{\int\! \d^2 \xi \sqrt{\det g} \, X_\m^2(\xi)}{DV}\right)   .
\eeq
The Gaussian action in $X$ implies that
\bea
\label{2.4}
\la X^2 \ra_V &=& \frac{1}{DV Z_V} \frac{\partial}{\partial \omega}
\int\! \cD [g]_V \cD [X_\m]_{\rm cm}\,\e^{-\!\int\! \d^2 \xi \sqrt{\det g}
  \, g^{ab} \prt_a X_\m \prt_b X_\m + \omega \!\int\! \d^2 \xi \sqrt{\det g}
  X_\m^2(\xi)}\Bigg\vert_{\,\omega=0} \nonumber\\
&=& \frac{1}{DV Z_V} \frac{\partial}{\partial \omega}
\int\! \cD [g]_V\Big({\rm det}' (-\Del_g-\omega)\Big)^{-\!D/2}\;
\Bigg\vert_{\,\omega=0} \nonumber\\
&=& \frac{1}{2 V Z_V} \int\! \cD[g]_V
\Big({\rm det}' (-\Del_g)\Big)^{-\!D/2}\;
\tr'\! \left[\frac{1}{-\Del_g}\right]\nonumber\\
& =& 
\frac{1}{2V}\la\tr'\!\left[\frac{1}{-\Del_g}\right]\ra_V,
\eea
where the primes on the determinants and traces again mean that zero modes are
excluded. Formula \rf{2.4} is used to define $\la X^2 \ra_V$ when $D$ is 
non-integer.
From \rf{2.2b} we obtain
\beq{2.5}
\la X^2 \ra_V  \sim V^{2/d_s -1},
\eeq
for $V$ going to infinity, and  from \rf{2.2}
we now conclude that 
\beq{2.7}
\frac{1}{d_s} = \frac{1}{D_H} +\frac{1}{2}.
\eeq
Since it is well-known  
that $D_H= \infty$ for $D$ Gaussian fields 
coupled to 2d quantum gravity \cite{kawai2}, as long as $D \leq 1$ we reach 
again the conclusion that $d_s=2$ for such systems.

Note that the scaling relation \rf{2.7} is valid for other 
self-similar systems like branched polymers. For the branched polymer 
systems we have $D_H=4$ and we are led to the known, but non-trivial 
result that $d_s=4/3$, the famous Alexander-Orbarch value, in the 
case of branched polymers \cite{jw}. Formula \rf{2.7} also 
applies for so-called multicritical 
branched polymers \cite{adj}, where $D_H= 2m/(m-1)$, $m \geq 2$.

\section{Lorentzian signature and causal structures}\label{causal}
As mentioned in the introduction, the solution of two-dimensional 
quantum gravity via an intermediate discretization
amounts to the counting of inequivalent geometries. 
While this counting problem 
has been solved in Euclidean gravity by the use of dynamical 
triangulations, an extension to
space-times with Lorentzian signature is not entirely
straightforward. 
We propose here a discretized model, formulated in the spirit 
of dynamical triangulations, where 
a causal structure -- characteristic for Lorentzian space-time
metrics --
is explicitly present in all the geometries 
included in the path integral.

The model is defined as follows. The topology of the underlying manifold
is taken to be $S^1\times [0,1]$, with ``space" represented by the closed 
manifold $S^1$. We consider the evolution of this space in ``time''. 
For the moment, no spatial topology changes are allowed, but we will 
return to this issue later.

\begin{figure}
\centerline{\hbox{\psfig{figure=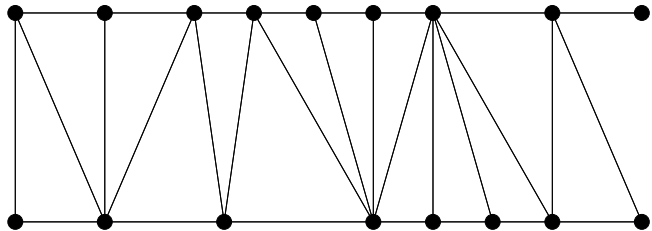,height=3cm,angle=0}}}
\caption[fig1]{The propagation of a spatial slice from step $t$ to 
step $t+1$. The ends of the strip should be joined to form a band
with topology $S^1 \times [0,1]$.}
\label{fig1}
\end{figure}

The geometry of each spatial slice is uniquely characterized by 
the length assigned to it. In the discretized version, the length $L$ 
will be quantized in units of a lattice spacing $a$, i.e.\ 
$L= l\cdot a$ where $l$ is an integer. A slice will thus be 
defined by $l$ vertices and $l$ links connecting them. 
To obtain a 2d geometry, we will evolve this spatial loop in 
discrete steps. This leads to a preferred notion of (discrete) ``time'' 
$t$, where each loop represents a slice of constant $t$.
The propagation from time-slice 
$t$ to time-slice $t+1$ is governed by the following rule: each vertex
$i$ at time $t$ is connected to $k_i$ vertices at time $t+1$, $k_i \geq 1$,
by links which are assigned length $-a$. The $k_i$ vertices, $k_i > 1$, 
at time-slice $t+1$ will be connected by $k_i-1$ consecutive 
space-like links, thus forming $k_i -1$ triangles. 
Finally the right boundary vertex
in the set of $k_i$ vertices will be identified with the left boundary 
vertex of the set of $k_{i+1}$ vertices. In this way we get a total of 
$\sum_{i=1}^l (k_i-1)$ vertices (and also links) at time-slice $t+1$ and 
the two spatial slices are connected by $\sum_{i=1}^l k_i
\equiv l_{t}+l_{t+1}$ triangles.
This is illustrated by fig.\ \ref{fig1}. 

The elementary building blocks of a geometry are therefore triangles
with one space- and two time-like edges. We define them to be flat
in the interior. A consistent way of assigning interior angles to
such Minkowskian triangles is described in \cite{sorkin}. The
angle between two time-like edges is $\gamma_{tt}=-\arccos \frac{3}{2}$,
and between a space- and a time-like edge $\gamma_{st}=
\frac{\pi}{2}+\frac{1}{2} \arccos \frac{3}{2}$, summing up to
$\gamma_{tt}+2\gamma_{st}=\pi$. The sum over all angles around a
vertex with $j$ incoming and $k$ outgoing time-like edges (by
definition $j,k\geq 1$) is given by $2\pi+(4-j-k)\arccos\frac{3}{2}$.
The regular triangulation of flat Minkowski space
corresponds to $j=k=2$ at all vertices. The volume of a single
triangle is given by $\frac{\sqrt{5}}{4}a^2$.

One may view these geometries as a subclass of all possible
triangulations that allow for the introduction of a causal
structure. Namely, if we think of all time-like links as being
future-directed, a vertex $v'$ lies in the future of a
vertex $v$ iff there is an oriented sequence of time-like
links leading from $v$ to $v'$. Two arbitrary vertices may or may not
be causally related in this way. 

In Lorentzian quantum gravity we are 
instructed to sum over all geometries connecting, 
say, two spatial boundaries of length $L_1$ and $L_2$, with the weight 
of each geometry $g$ given by 
\beq{1}
\e^{i S[g]}, ~~~~~S[g] = \L \int \!\!\sqrt{-\det g}~~~(\mbox{in 2d}),
\eeq
where $\L$ is the cosmological constant.
In our discretized model the boundaries will be characterized by 
integers $l_1$ and $l_2$, the number of vertices or links at the two
boundaries. The path integral amplitude for the propagation from 
geometry $l_1$ to $l_2$ will be the sum over all interpolating 
surfaces of the 
kind described above, with a weight given by the discretized version of 
\rf{1}. Let us call the corresponding amplitude $G_\l(l_1,l_2)$,
which may be represented as the sum
\beq{new1}
G_\l(l_1,l_2) = \sum_{t=1}^{\infty} G_\l(l_1,l_2;t).
\eeq

As already mentioned, $G_\l(l_1,l_2;1)$ serves as a transfer matrix
for the model.
Knowledge of $G_\l(l_1,l_2;1)$ allows us to find $G_\l(l_1,l_2;t)$
by iterating \rf{8} $t$ times. 
Using standard techniques 
of generating functions we associate a factor $g=e^{i\l}$ 
with each 
triangle (not to be confused with the metric $g_{ab}$), 
a factor $x$ with each vertex on the entrance loop and 
a factor $y$ with each vertex on the exit loop, leading to
\beq{12}
G_\l(x,y;1) =\sum_{k=0}^\infty \left( gx \sum_{l=0}^\infty
 (gy)^l \right)^k -
\sum_{k=0}^\infty (gx)^k = \frac{g^2 xy}{(1-gx)(1-gx-gy)}.
\eeq
Formula \rf{12} is simply a book-keeping device for all possible
ways of evolving from an entrance loop of any length in one step to
an exit loop of any length.
The subtraction of the term $1/(1-gx)$ has been performed to 
exclude the degenerate cases where either the entrance or the exit 
loop is of length zero. 
  
From \rf{12} and eq.\ \rf{10}, with $t_1=1$, we obtain
\beq{13}
G_\l(x,y;t) = \frac{gx}{1-gx}\; G_\l(\frac{g}{1-gx},y;t-1).
\eeq
This equation can be iterated and the solution written as 
\beq{14}
G_\l(x,y;t) = F_1^2(x)F_2^2(x) \cdots F_{t-1}^2(x) 
\frac{g^2 xy}{[1-gF_{t-1}(x)][1-gF_{t-1}(x)-gy]},
\eeq
where $F_t(x)$ is defined iteratively by
\beq{15}
F_t(x) = \frac{g}{1-gF_{t-1}(x)},~~~F_0(x)=x.
\eeq
Let $F$ denote the fixed point of this iterative equation. By standard
techniques one readily obtains
\beq{16}
F_t(x)= F\ \frac{1-xF +F^{2t-1}(x-F)}{1-xF +F^{2t+1}(x-F)},~~~~
F=\frac{1-\sqrt{1-4g^2}}{2g}.
\eeq
Inserting \rf{16} in eq.\ \rf{14}, we can write
\bea
G_\l(x,y;t)\!\!\! &=&\!\!\!\!  \frac{ F^{2t}(1-F^2)^2\; xy}
{(A_t-B_tx)(A_t-B_t(x+y)+C_txy)}
\label{17}\\
~\!\!\!& =&\!\!\!\!
 \frac{F^{2t}(1-F^2)^2\;xy}{\Big[(1\!\!-\!xF)\!-\!F^{2t+1}(F\!\!-\!x)\Big]
\Big[(1\!\!-\!xF)(1\!\!-\!yF)\!-\!F^{2t} (F\!\!-\!x)(F\!\!-\!y)\Big]}~,
~~~\label{17a}
\eea
where the time-dependent coefficients are given by 
\beq{18}
A_t =1-F^{2t+2},~~~B_t=F(1-F^{2t}),~~~C_t=F^2(1-F^{2t-2}).
\eeq
The combined region of convergence of the 
expansion in powers $g^kx^ly^m$, valid for all $t$, is 
\beq{18a}
|g| < \oh,~~~~ |x|< 1,~~~~|y|<1.
\eeq

We have analyzed the possible continuum limits that
may be obtained for expression \rf{17} (see \cite{al} for details).
The result is that one only gets an interesting continuum limit 
if the divergent (unphysical) parts of 
the bare cosmological constants are taken to be {\it purely imaginary}.
The Lorentzian form for the continuum 
propagator is then obtained by an analytic continuation $\L\to -i\L$
in the {\it renormalized} coupling of the resulting Euclidean 
expressions. 
At this stage it may seem that we are surreptitiously reverting
to a fully Euclidean model. We could of course equivalently 
have conducted
the entire discussion up to this point in the ``Euclidean sector'',
by omitting the factor of $-i$ in the exponential \rf{1} of the
action, choosing $\l$ positive real and taking all edge lengths 
equal to 1. However, from a purely Euclidean point of view {\it there
is no reason for restricting the state sum to a subclass
of geometries admitting a causal structure}. As we will show
below, {\it this restriction changes the quantum theory drastically}. 
The associated 
preferred notion of a discrete ``time'' allows us to define
an ``analytic continuation in time''. Because of the simple
form of the action in two dimensions, the rotation 
\beq{25b}
\int  dx\ dt \sqrt{-\det g_{lor}} \to i\int  dx \ dt_{eu} 
\sqrt{\det g_{eu}}
\eeq
to Euclidean metrics in our model is equivalent to the analytic continuation
of the cosmological constant $\L$.

Rather than discussing in detail how to obtain a continuum limit from 
\rf{17}, let us show how one can derive a differential equation
for the continuum limit $G_\L(X,Y;T)$ of $G_\l(x,y;t)$.
Inserting \rf{*l3a} 
in eq.\ \rf{13}
and expanding to first order in the lattice spacing $a$ leads to
\beq{32} 
a^{\ep-1}\frac{\prt}{\prt T} G_\L(X,Y;T) + \frac{\prt}{\prt X}
\Bigl[ (X^2-\L) G_\L(X,Y;T) \bigr]=0.
\eeq
We conclude that an interesting continuum 
limit exists {\it provided the critical exponent $\ep$ in $T = t a^{\ep}$
is equal to 1}. This value is different from 
the one obtained for Euclidean gravity where $\ep =1/2$.
With $\ep =1$, equation \rf{32} becomes
a standard first-order partial differential equation which 
should be solved subject to the boundary condition 
\beq{28}
G_\L(X,Y;T\equ 0) = \frac{1}{X+Y},
\eeq
which expresses the condition $G_\L(L_1,L_2;T\equ 0)=\del(L_1\mi L_2)$.
The solution is 
\bea
G_\L(X,Y;T) &=& \frac{4\L\ \e^{-2\SLT}}{(\SL+X)+\e^{-2\SLT}(\SL-X)}\nonumber\\
&&\times \, \frac{1}{(\SL+X)(\SL+Y)-\e^{-2\SLT}(\SL-X)(\SL-Y)},
\label{26}
\eea
or, after an inverse Laplace transformation,
\beq{30}
G_\L(L_1,L_2;T) = \frac{\e^{-[\coth \SLT] \SL(L_1+L_2)}}{\sinh \SLT}
\; \frac{\sqrt{\L L_1 L_2}}{L_2}\; \; 
I_1\left(\frac{2\sqrt{\L L_1 L_2}}{\sinh \SLT}\right), 
\eeq
where $I_1(x)$ is a modified Bessel function of the first kind.
Remarkably, our highly non-trivial 
expression \rf{30} agrees
with the loop propagator obtained from a bona-fide continuum calculation
in proper-time gauge of pure 2d gravity by Nakayama \cite{nakayama}.

This result is clearly different from the 
one obtained in Euclidean gravity. 
The ``proper time" $T$ has the dimension of $1/\SL$, 
while the dimensionality of the analogous quantity in
Euclidean gravity, the geodesic 
distance $R$, is that of $1/\sqrt[4]{\L}$.
This also implies that for large $T$ the average 
``spatial'' volume of a slice at some intermediate time will
have an anomalous dimension. In fact, in Euclidean 2d quantum gravity  
we have 
\beq{ob7}
G^{(eu)}_\L(L_1,L_2;R) \propto \e^{-\sqrt[4]{\L} R}~~~~\mbox{for}~~~
R \to \infty,
\eeq
from which one can calculate the average {\em two-dimensional} 
volume $V(R)$ in 
the ensemble of universes with two boundaries separated by a
geodesic distance $R$,
\beq{ob8}
\la V(R) \ra = - \frac{1}{G^{(eu)}_\L(L_1,L_2;R)} \frac{\d}{\d \L}\;
G^{(eu)}_\L(L_1,L_2;R) \propto \frac{R}{\L^{3/4}}.
\eeq
For large $R$ we therefore expect the average spatial volume $L_{space}$ 
at intermediate $T$'s to behave like
\beq{ob9}
\la L_{space} \ra = \frac{\la V(R) \ra}{R} \propto \frac{1}{\L^{3/4}}.
\eeq
In the present model, according to \rf{26}, the amplitude
behaves for large $T$ like 
\beq{ob10}
G_\L(L_1,L_2;T) \propto \e^{-\SL T},
\eeq
which implies that the dimension of $T$ in this case is 
$[{\rm L}]$. Instead of \rf{ob9}, we therefore obtain the dependence
\beq{ob11}
\la L_{space} \ra \propto \frac{1}{\SL}.
\eeq
This reflects the fact that the quantum space-time of the model does 
not have an anomalous fractal dimension, and thus differs drastically 
from the average space-time in the usual theory of
two-dimensional Euclidean quantum 
gravity. Recall that in the classification of the critical behaviour in terms 
of $(\ep,\eta)$, for purely ``kinematical reasons'', illustrated
in fig.\ref{identity}, we needed either $(\ep,\eta)=(1/2,3/2)$ (Euclidean 
quantum gravity) or $\ep=1, \eta <0$. It can be shown that 
in the present model $\eta = -1$ (see \cite{al} for details).
We do not know if alternative models with different 
$\eta$ exist.

\subsection{Topology changes of space}

So far in our non-perturbative regularization of {\it Lorentzian} 
2d quantum gravity we 
have not considered the possibility of spatial topology changes.
We will now show that {\it if} one includes such
topology changes one recovers the usual scaling of
Euclidean 2d quantum gravity.
By a topology change of {\em space} in our Lorentzian setting
we have in mind the following: a baby universe may
branch off at some time $T$ and develop in the future, where it 
will eventually disappear into the vacuum. However, it is
not allowed to rejoin the ``parent'' universe and thus change the 
overall topology of the two-dimensional manifold. This is 
a restriction we impose to be able to compare with 
the analogous calculation in usual 2d Euclidean quantum gravity.

\begin{figure}
\centerline{\hbox{\psfig{figure=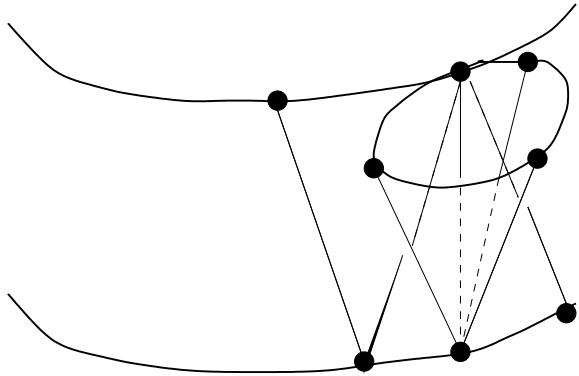,height=5cm,angle=0}}}
\caption[branching]{A ``baby universe'' branches off locally in 
one time-step.}
\label{branching}
\end{figure}

\begin{figure}
\centerline{\hbox{\psfig{figure=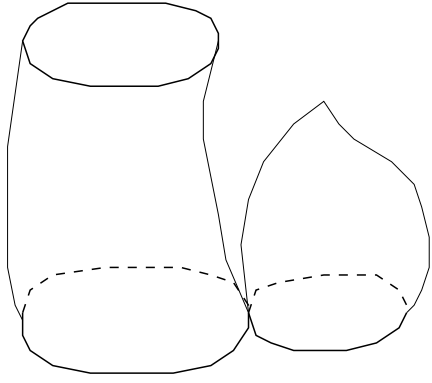,height=5cm,angle=0}}}
\caption[topchange]{A ``baby universe'' is created by a global pinching.}
\label{topchange}
\end{figure}

Our starting point will be the discretized model introduced above.
The superscript $^{(b)}$ will indicate the 
``bare'' model without spatial topology changes.
There are a number of ways to implement the creation of baby universes, 
some more na\-tu\-ral than others. They all agree in the continuum limit,
as will be clear from the general arguments provided below. 
We mention just two ways of implementing such a change. The first
is a simple generalization of the forward step we have used
in the original model, where each vertex at time $t$ could connect to $n$
vertices at time $t+1$. We now allow in addition that these 
sets of $n$ vertices (for $n >2$) may form a baby universe with
closed spatial topology $S^1$, branching off from the rest.
The process is illustrated in fig.\ \ref{branching}.
An alternative and technically somewhat simpler way to implement the 
topology 
change is shown in fig.\ \ref{topchange}: stepping forward from 
$t$ to $t+1$ from a loop of length $l_1$ we create a
baby universe of length $l < l_1$ by pinching it off non-locally
from the main branch. We have checked that the 
continuum limit is the same in both cases. For simplicity we only 
present the derivation in the latter case.

Accounting for the new possibilities of evolution in each step
according to fig.\ \ref{topchange}, the new and old transfer matrices
are related by 
\beq{top1}
G_\l(l_1,l_2;1) = G_\l^{(b)}(l_1,l_2;1)+ 
\sum_{l=1}^{l_1-1} l_1 w(l_1\mi l,g)\,G_\l^{(b)}(l,l_2;1).
\eeq
The factor $l_1$ in the sum comes from the fact that the 
``pinching'' shown in fig.\ \ref{topchange} can take place at any of the 
$l_1$ vertices. As before, the new transfer matrix leads to new amplitudes
$G_\l(l_1,l_2;t)$, satisfying
\beq{top2a}
G_\l(l_1,l_2;t_1+t_2) = \sum_l G_\l(l_1,l;t_1)G_\l(l,l_2;t_2),
\eeq
and in particular 
\beq{top2}
G_\l(l_1,l_2;t) = \sum_l G_\l(l_1,l;1)\;G_\l(l,l_2;t\mi 1).
\eeq
Performing a (discrete) Laplace transformation of eq.\ \rf{top2}
leads to
\beq{top3}
G(x,y;g;t) = 
\ointz \left[G_\l^{(b)}(x,z^{-1};1)+ x \frac{\prt}{\prt x} 
\Bigl( w(x;g) G_\l^{(b)}(x,z^{-1};1)\Bigr) \right]  G(z,y;g;t \mi 1), 
\eeq 
or, using the explicit form of the transfer matrix $G_\l^{(b)}(x,z;1)$,
formula \rf{12},
\beq{top4}
G(x,y;g;t) = \Bigl[1+x\frac{\prt w(x,g)}{\prt x}
+ x w(x,g)\frac{\prt}{\prt x} \Bigr]
\, \frac{gx}{1-gx} \, G\Bigl( \frac{g}{1\mi gx},y;g;t\mi 1\Bigr).
\eeq

Proceeding as before, we can now derive a differential equation 
in the continuum limit. We will assume that the disc amplitude
scales according to \rf{an3}, leading to
\bea
a^\ep\frac{\prt}{\prt T}\, G_\L(X,Y;T)& =&
-a \, \frac{\prt}{\prt X} \Bigl[ (X^2-\L) G_\L(X,Y;T)\Bigr] \nn
&&-a^{\eta-1}\frac{\prt}{\prt X} \Bigl[W_\L(X) G_\L(X,Y;T)\Bigr].
\label{top13}
\eea
The first term on the right-hand side of eq.\ \rf{top13} 
is precisely the one we have already encountered 
in the model without baby universe creation, while the second term 
is due to the creation of baby universes. 
Clearly the case $\eta < 0$ (in fact $\eta \leq 1$) is inconsistent
with the presence of the second term, i.e.\ the creation of baby universes. 
However, since $\eta <2$, 
the last term on the right-hand side of \rf{top13} always dominates
the first term. {\em Once we 
allow for the creation of baby universes, this process will completely
dominate the continuum limit.} In addition we get $\ep = \eta-1$,
in agreement with relation \rf{an6}. Thus $\eta > 1$ and 
we conclude that $\ep=1/2,\eta=3/2$ 
are the only possible scaling exponents if we allow for the creation 
of baby universes. 
These are precisely the scaling exponents obtained from 
two-dimensional Euclidean gravity in
terms of dynamical triangulations. The topology 
changes of space have induced an anomalous dimension for $T$. 
If the second term on the right-hand side of \rf{top13} had been absent,
this would have led to $\ep =1$, and the time $T$ scaling in the 
same way as the spatial length $L$. --
In summary, eq.\ \rf{top13} leads to the continuum relation
\beq{top16}
\frac{\prt}{\prt T}\, G_\L(X,Y;T) =
-\frac{\prt}{\prt X} \Bigl[W_\L(X) G_\L(X,Y;T)\Bigr],
\eeq
which coincides with  an analogous equation that can be derived directly in
2d Euclidean quantum gravity.

\section{Discussion}

In the present article we have demonstrated by explicit calculation
that it {\it is} possible to formulate a background-independent theory 
of quantum gravity and still maintain some analogues of
concepts from classical gravity, like ``rods-and-clocks'' measurements
of the (quantum) geometry, and propagating
test particles to probe the (quantum) geometry.
Clearly it is a major limitation that all calculations are 
performed in two dimensions, and until recently have
been limited to the Euclidean sector of quantum gravity.
However, {\it any} theory of quantum gravity will 
have to address questions similar the ones we have been dealing
with in the context of two-dimensional gravity.

Our results were two-fold: first it was shown that the generic
two-dimensional geometry is {\it fractal}, with intrinsic Hausdorff 
dimension four, and not two. This si\-tu\-ation is reminiscent 
of the random walk  
representation for the free particle which is of immense importance 
for our understanding of relativistic quantum field theory.
The fact that the fractal dimension of a generic walk 
is two and not one, as one would naively have expected, 
underscores the impossibility of associating 
a classical path with a particle at distances smaller than its inverse
mass. However, the concept of intrinsic dimensionality 
is more complex in two-dimensional quantum gravity, as exemplified by 
our study of the propagation of test particles. The dimensionality
obtained from the return probability of a diffusion process is 
{\it two}, and not four. Ne\-ver\-theless, the actual diffusion process
{\it is} anomalous, in the sense that the average geodesic distance 
travelled grows only like $\tau^{1/4}$, and not $\tau^{1/2}$,
as a function of the fictitious diffusion time $\tau$, 
as would be the case on any non-fractal manifold.
However, this anomalous propagation behaviour is counter-balanced by the 
anomalous volume contained within a radius of $\tau^{1/4}$,
so that in 
the end the return probability is the same as on a two-dimensional smooth 
manifold.

Secondly, we have discussed the relation between Euclidean quantum 
gravity, defined via a non-perturbative path integral, and 
a similar construction in the Lorentzian case. Again by 
explicit calculation, using a simple discretized Lorentzian model 
where causal structures were included in a natural way, we
obtained a conti\-nu\-um limit different from that
of Euclidean quantum gravity, but which can again be 
reproduced by formal continuum calculations. 
The reason 
for the difference could be traced to the presence or absence of 
topology changes of space. In standard classical Lorentzian 
gravity such topology changes are associated with singular points
of the metric structure, but none of the usual 
principles of quantization tell us whether such configurations
should be allowed quantum-mechanically or not.
However, we showed that we can only obtain agreement with 
the theory of Euclidean quantum gravity if we allow for spatial topology
changes.

This state of affairs is perfectly acceptable from a physical
point of view: 
unlike in the Lorentzian case, in Euclidean quantum gravity we have 
no distinction between space- and time-like directions. 
If we consider the cylinder 
amplitude, some choices of ``time''-slicing may result in baby universe 
creation as a function of ``time'', while others may not. 
The concept of ``absence of baby universes" makes only
sense in a situation where there is a (class of) preferred time
choice(s), as was the case for our discrete Lorentzian geometries.
In the Lorentzian case, we had to explicitly allow for the creation of 
baby universes to obtain a class of geometries matching
those summed over in Euclidean quantum gravity. This choice 
resulted then (up to analytic continuation) in the already known
Euclidean quantum theory. 
Similar intricacies in the relation between Euclidean and Lorentzian
quantum gravity are to be expected also in higher dimensions.

\section*{Acknowledgement}
J.A. acknowledges the support from MaPhySto which is financed
by the Danish National Research Foundation.


\begin{thebibliography}{77}

\def \np {{\it Nucl. Phys. }}
\def \pl {{\it Phys. Lett. }}


\def \mpl {{\it Mod. Phys. Lett. }}

\bibitem{david1}F. David, \np {\bf B257} (1985) 45;
 \np {\bf B257} (1985) 543;
\bibitem{adf}J. Ambj\o rn, B. Durhuus and J. Fr\"{o}hlich,
 \np {\bf B257} (1985) 433; 
\bibitem{kazakov}V.A. Kazakov, Phys.Lett.\ 150B (1985) 282.
\bibitem{kkm}V.A. Kazakov, I.K. Kostov and A.A. Migdal, 
\pl {\bf 157B} (1985) 295.
\bibitem{watabiki}Y. Watabiki, Prog.Theor.Phys.Suppl. 114
(1993) 1.
\bibitem{japanese}H.\ Kawai, N.\ Kawamoto, T.\ Mogami and Y.\ Watabiki,
{Phys.\ Lett.\ B}\ 306 (1993) 19-26.
\bibitem{sorkin} R.\ Sorkin, 
{ Phys.\ Rev.\ D}\ 12 (1975) 385-396; Err. { ibid.}\ 23 (1981) 565.
\bibitem{jw}T. Jonsson and J.F. Wheater, Nucl.Phys. B515 (1998) 549;\\
J.D. Correia and J.F. Wheater, Phys.Lett. B422 (1998) 76.
\bibitem{al} J.\ Ambj\o rn and R. Loll, {\it 
Non-perturbative Lorentzian Quantum Gravity, Causa\-lity and Topology Change},
hep-th/9805108.
\bibitem{abnrw}J. Ambjorn, D. Boulatov, J.L. Nielsen, J. Rolf, Y. Watabiki,
J.High Energy Phys. 02 (1998) 010.
\bibitem{nakayama}R.\ Nakayama, 
{ Phys.\ Lett.\ B}\ 325 (1994) 347-353. 
\bibitem{adfo}J. Ambj\o rn, B. Durhuus  J. Fr\"{o}hlich and P. Orland,
 \np {\bf B270} (1986) 457;
\bibitem{david}F. David, \mpl {\bf A5} (1990) 1019.
\bibitem{ajm}J. Ambj{\o}rn, J. Jurkiewicz and Y. M. Makeenko,
\pl {\bf B251} (1990) 517.
\bibitem{aw}J. Ambj\o rn and Y. Watabiki,
\np {\bf B445} (1995) 129. 
\bibitem{ajw}J. Ambj\o rn, J. Jurkiewicz and Y. Watabiki,
\np {\bf B454} (1995) 313.
\bibitem{review}S.\ Havlin and D.\ Ben-Avraham, {\it Adv.\ Phys.} {\bf 36},
(1987) 695. \bibitem{adj}J.\ Ambj\o rn, B.\ Durhuus and T.\ Jonsson, 
Phys.Lett.B244 (1990) 403.
\bibitem{book1}J.\ Ambj\o rn, B.\ Durhuus and T.\ Jonsson, {\it Quantum
Geometry}, Cambridge Monographs on Mathematical Physics, Cambridge, 1997.
\bibitem{kawai2} H. Kawai, {\it Nucl.\ Phys.\ Proc.\ Suppl.} {\bf 26} 
(1992) 93.

\end{thebibliography}
\end{document}